\documentclass[a4paper]{spie} 

\usepackage{graphicx}
\usepackage{url}
\usepackage[latin1]{inputenc}
\usepackage[OT1]{fontenc}

\usepackage{graphicx}

\newcommand{\dg}{$^\circ$}

\newcommand{\brg}{Br$_{\gamma}$}
\newcommand{\zcma}{Z\,CMa}

\begin{document}

\title{The 2008-2009 outburst of the young binary system Z CMa
  unraveled by interferometry with high spectral resolution} 

\author{%
 Fabien Malbet\supit{a}, 
 Myriam Benisty\supit{b},
 Catherine Dougados\supit{a},
 Antonella Natta\supit{b}, \\
 Jean-Baptiste Le Bouquin\supit{a}, 
 Fabrizio Massi\supit{b}, 
 J\'er\^ome Bouvier\supit{a},
 Konstantin Grankin\supit{c}, \\
 Mickael Bonnefoy\supit{a}, 
 Emma Whelan\supit{a}
 \skiplinehalf 
 \supit{a} Laboratoire d'Astrophysique de Grenoble (LAOG), UMR 5571
 Université Joseph Fourier/CNRS, BP 53, F-38051 Grenoble cedex 9,
 France;\\
 \supit{b} INAF-Osservatorio Astrofisico di Arcetri, Largo
 E.~Fermi 5, 50125 Firenze, Italy\\
 \supit{c} Crimean Astrophysical Observatory, 98409 Nauchny, Crimea,
 Ukraine
}%


 
\maketitle 
\begin{abstract}
 Z CMa is a young binary system consisting of an Herbig primary and a
 FU Ori companion. Both components seem to be surrounded by active
 accretion disks and a jet was associated to the Herbig B0. In
 Nov. 2008, K.\ Grankin discovered that Z CMa was exhibiting an
 outburst with an amplitude larger than any photometric variations
 recorded in the last 25 years. To study the innermost regions in
 which the outburst occurs and understand its origin, we have
 observed both binary components with AMBER/VLTI across the \brg\
 emission line in Dec. 2009 in medium and high spectral resolution modes. Our observations show
 that the Herbig Be, responsible for the increase of luminosity, also
 produces a strong \brg\ emission, and they allow us to disentangle
 from various origins by locating the emission at each velocities
 through the line. Considering a model of a Keplerian disk alone
 fails at reproducing the asymmetric spectro-astrometric
 measurements, suggesting a major contribution from an outflow.
\end{abstract}



\section{Introduction}

Accretion plays an important role in star and planet formation. For
many years, it was considered to be a slow quasi-stationary
process\cite{stahler98}, occurring mostly through a viscous disk
ending in its inner part by a boundary layer\cite{bertout88} with the
star or by magnetospheric funnels\cite{konigl91,
 calvet92}. However, this scenario has been challenged by
observations\cite{kenyon90, evans09} that suggest that the
accretion process could be time-variable and occur quickly by means of
short high mass accretion rate bursts. Studying the very inner region
of a young stellar object that is known to experience episodic
photometric outbursts is thus of prime importance to understand the
role of accretion in the formation of the star and its environment.

The study presented in this paper is part of a large observational
campaign targeting \zcma\ during its 2008 outburst that aims to
understand its origin. To directly probe the morphology of the hot gas
in the inner AUs, we took advantage of the spatial and spectral
resolution available at the VLTI to perform micro-arcsecond
spectro-astrometry. With AMBER, we resolve the $K$-band emission of
the hot gas surrounding \emph{each} star at the milliarcsecond
resolution. This paper outlines the analysis already published by
Benisty et al. (2010)\cite{2010arXiv1007.0682B}.

Section \ref{sec:zcma} summarizes what we know about \zcma.  In
Sect.~\ref{sec:obs}, we present the observations and the data
processing. Section \ref{sec:results} presents the findings of our
study that we discuss in Sect.~\ref{sec:discussion}.

\section{The system Z Canis Majoris}
\label{sec:zcma}

\begin{figure}[t]
  \centering
  \includegraphics[width=0.95\hsize]{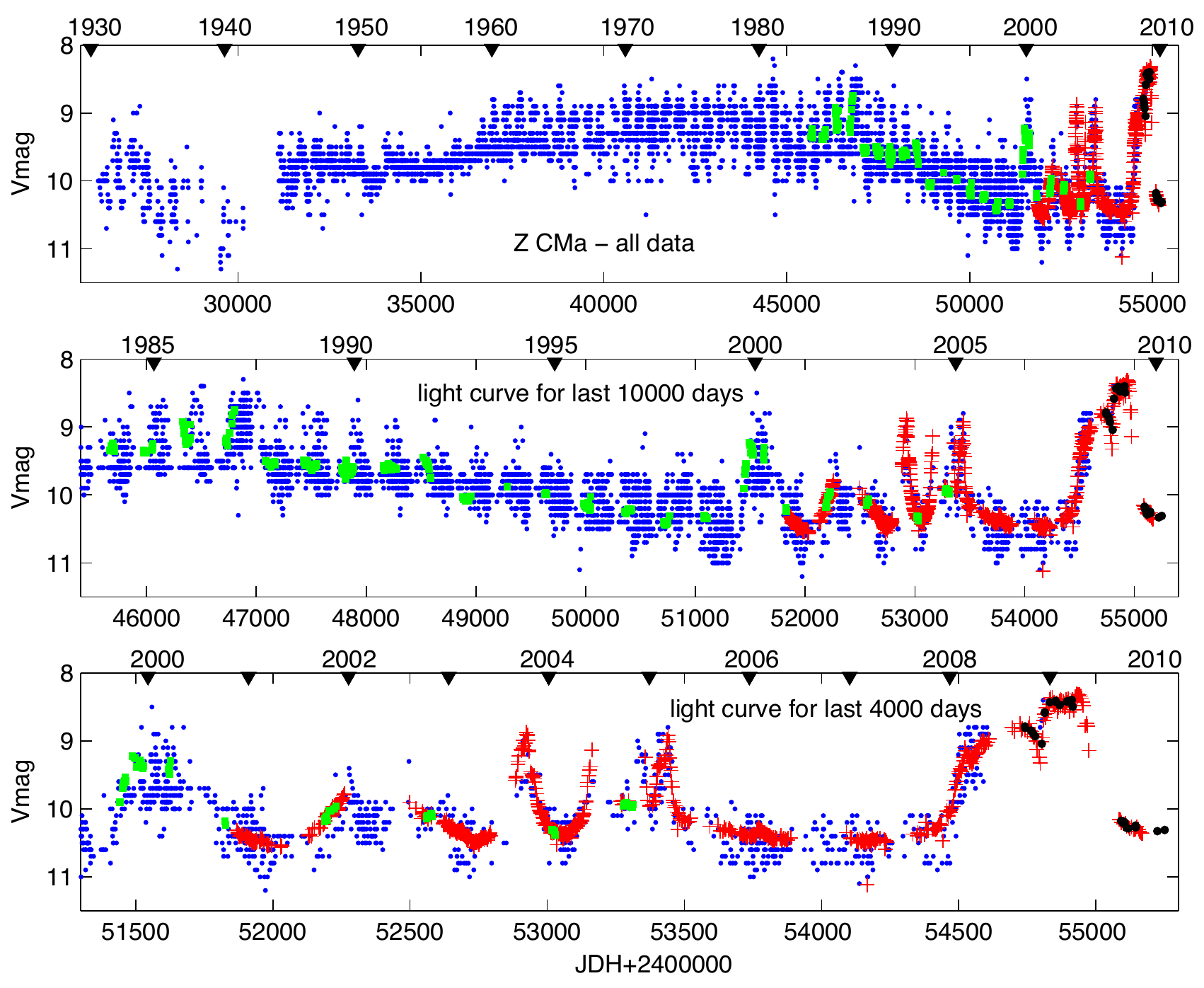}
  \caption{Light curve of Z CMa showing the 2008 outburst (see Grankin
    et al.\ 2009\cite{grankin09}) complemented with the last data. Top
    panel: all data observed since 1930. Middle panel: the last 25
    years. Bottom panel: the last ten years showing the sudden raise
    of $V$ by two magnitudes in 2008 and which disappears beginning of
    2010. Black filled circles: photometric observations from CrAO;
    green squares: observations from Mt.\ Maidanak; red crosses: CCD
    observations from ASAS; and blue points: visual estimations of a
    brightness from AAVSO. }
  \label{fig:photometry}
\end{figure}

\zcma\ is a pre-main-sequence binary with a separation\cite{koresko91,
  barth94} of 0.1'' located at a distance estimated\cite{claria74,
  kaltcheva00} from 930 to 1150\,pc. The primary, embedded in a dust
cocoon, was identified as a Herbig~Be star based on
spectropolarimetry\cite{whitney93}. It is surrounded by an inclined
disk, possibly a circumbinary disk, as inferred from millimeter
observations\cite{alonso09}, and dominates the infrared continuum and
total luminosity of the system. In contrast, the secondary is the
major source of continuum emission at visual wavelengths. Although the
secondary has not undergone a large outburst this century, it was
identified as a FU~Or object based on its broad double-peaked optical
absorption lines, which are typical of a circumstellar disk that
undergoes a strong accretion, and spectral type\cite{hartmann89} of
F-G. In the past twenty years, the \zcma\ system exhibited repeated
brightness variations, of $\sim$0.5-1 visual magnitude, which were
attributed to the Herbig~Be star\cite{vandenancker04}. \zcma\ is
clearly associated with a bipolar outflow\cite{poetzel89, evans94}
that extends to 3.6\,pc along PA$\sim$240\dg. A 1"x0.24"
micro-jet\cite{garcia99} was detected in the [OI]~6300\AA~line in the
same direction, and the authors of this work\cite{garcia99} concluded
that the optical emission-line spectrum and the jet are associated
with the primary. However, the innermost environments of the \zcma\
components have been poorly studied. Two broad-band interferometric
measurements have been obtained, allowing only characteristic
sizes\cite{monnier05,millan06} of the $K$-band continuum emission to
be derived.

In January 2008, \zcma's brightness increased\cite{grankin09} by about
two visual magnitudes, representing the largest outburst observed in
the past 90 years (see Fig.~\ref{fig:photometry}).  Based on spectropolarimetric
observations\cite{szeifert10}, this outburst was considered to be
associated with the Herbig~Be star. The overall spectral energy distribution of the
system is strongly modified during the outburst at wavelengths shorter
than 10\,$\mu$m, which indicates that the outburst originates close to
the star.

\section{Observations and data processing}
\label{sec:obs}

\zcma\ was observed at the Very Large Telescope
Interferometer\cite{vlti1} (VLTI), using the AMBER
instrument\cite{petrov07} that allows the simultaneous combination of
three beams in the near-infrared. The instrument delivers spectrally
dispersed interferometric observables (visibilities, closure phases,
differential phases) at spectral resolutions up to 12\,000.

\begin{table}[t]
\centering
\caption{Log  of  the  observations.} 
\medskip
\label{tab:obs}
\begin{tabular}{cccccc}
 \hline
Date & Baseline & Projected & Position & $R$ & \\ 
& & length (m)& angle ($^\circ$)& & \\
 \hline
 \hline
 05/12/08 & D0-G1 & 69 & 137 & 1500 & FUOr+HBe \\
 07/12/08 & K0-G1 & 89 & 28 & 12000 & FUOr+HBe \\
 09/12/08 & K0-G1 & 88 & 24& 12000 & FUOr+HBe \\
 \hline 
 15/12/08 & U2-U3 & 44 & 35& 1500 & \\
 & U3-U4 & 62 & 107 & & \\
 & U2-U4 & 86 & 78 & & \\
 16/12/08 & U1-U2 & 56& 35& 1500 & \\
 & U2-U4 & 77& 88& & \\
 & U1-U4 &120 & 66& & \\
 \hline
 10/01/10 & D0-G1 & 71 & 133& 1500 & FUOr+HBe \\
 \hline
\multicolumn{6}{l}{\footnotesize Note: $R$  is  the  spectral
 resolution. 'FUOr+HBe' specifies when the binary is in the field of view.}
\end{tabular}
\end{table}
We report $K$-band observations taken in the medium
spectral resolution mode (MR; R$\sim$1500) with the 8.2\,m Unit
Telescopes (UTs) as well as with the 1.8\,m Auxiliary Telescopes (ATs),
and in the high spectral resolution mode (HR; R$\sim$12\,000) with the
ATs. The data were obtained within programs of Guaranteed Time,
Director's Discretionary Time, and Open Time observations. \zcma\ was
observed with 11 different baselines of 4 VLTI configurations, during
5~nights in December 2008 and one night in January 2010 (see
Table~\ref{tab:obs} for the summary of observations; the baselines
G1-A0 and K0-A0 which could not be used are not reported). The longest
baseline is $\sim$120\,m corresponding to a maximum angular resolution
of 3.7\,mas. With the UTs, the observations are
coupled with the use of adaptive optics and the resulting field of
view ranges from 50 to 60\,mas. This allowed us to spatially resolve
the binary and obtain \textit{separate} measurements of the FU~Or and
the Herbig~Be. In contrast, the ATs field-of-view, ranging from 230
to 280\,mas, includes both stars and the interferometric signal results
from both emissions. In addition to \zcma, calibrators (HD45420,
HD60742, HD55137, HD55832) were observed to correct for instrumental
effects. All observations were performed using the fringe-tracker
FINITO\cite{lebouquin08}. 

The data reduction was performed following the standard
procedures\cite{tatulli07,chelli09}, using the \texttt{amdlib}
package, release 2.99, and the \texttt{yorick} interface provided by
the Jean-Marie Mariotti
Center\footnote{\texttt{http://www.jmmc.fr/amberdrs}}.  Raw spectral
visibilities, differential phases, and closure phases were extracted
for all the frames of each observing file. A selection of 80\% of the
highest quality frames was made and to avoid the effects of
instrumental jitter and unsatisfactory light injection. Consecutive
observations were merged to enhance the signal-to-noise
ratio. Calibration of the AMBER+VLTI instrumental transfer function
was done using measurements of the calibrators, after correcting for
their diameter.  The accuracy of the wavelength/velocity calibration
is $\sim50$\,km/s.  Because the $K$-band continuum measured by the ATs
(due to both stars) is very resolved on long baselines ($V^2\sim0$), the
observations obtained on the G1-A0 and K0-A0 baselines could
unfortunately not be exploited. The absolute value of the visibilities
obtained with the UT baselines could not be determined due to random
vibrations of the telescopes. However, this issue affects all spectral
channels in the same way, and does not modify our conclusions.

\section{Findings}
\label{sec:results}

We recall that the visibilities provide information about the spatial
extent of the emission, and decrease as the extension increases.
Differential phases provide a measurement of the photocenter
displacements across the sky, projected along the baseline
direction. They can therefore be converted into differential
spectro-astrometric shifts. They are measured relative to the
continuum, for which we assume a zero phase.  Finally, the closure
phases are related to the asymmetry of the brightness distribution
(\textit{e.g.}, they are null for a point-symmetric object).

\subsection{The \brg\ line is present only during the outburst}
\label{sec:brg-outburst}

\begin{figure}[t]
 \centering
 \includegraphics[width=0.59\textwidth]{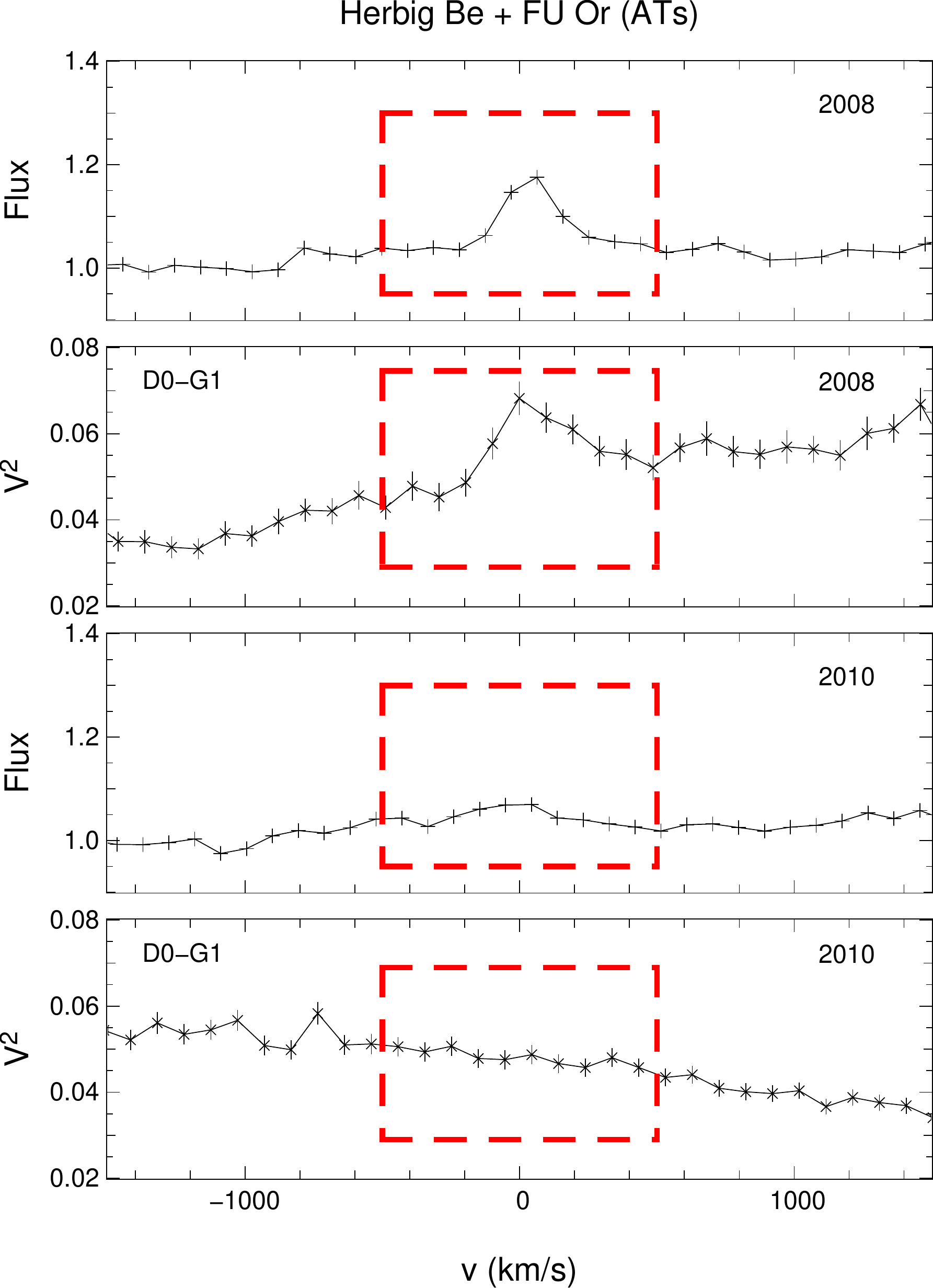} 
 \caption{\label{fig:brg-emission} Medium spectral resolution normalized
   spectra and squared visibilities during (2008) and after the
   outburst (2010). The slopes of the visibility curves depend on the
   binary characteristics and on the observing set up, that differs in
   the two observations. The field of view of the ATs contains both
   components of the binary system.  The \brg\ line is present only
   during the outburst.  Note the disappearance of the \brg\ line
   signature in the spectra and the visibility after the outburst (red
   dashed squares). } 
\end{figure}

Fig.~\ref{fig:brg-emission} compares the spectra and the
visibilities obtained during and after the outburst: the emission
line, and the signature in the visibilities, disappear after the
outburst. Plotted within a large velocity range, the visibilities show
a typical signature of binarity (\textit{i.e.}, a cosine modulation),
in agreement with the system main characteristics 
(separation, position angle, flux ratio; Bonnefoy et al., in prep.). 

The main consequence of the outburst at least for interferometric
observations is to produce a \brg\ in the spectrum and a raise of the
visibility which disappears when the outburst is over. Therefore, the
origin of the outburst must be searched in the emission line, the
\brg\ line being well-suited.

\subsection{The outburst is seen only in the Herbig Be component}
\label{sec:outburst-HBe}

\begin{figure}[t]
  \centering
  \includegraphics[height=0.65\textheight]{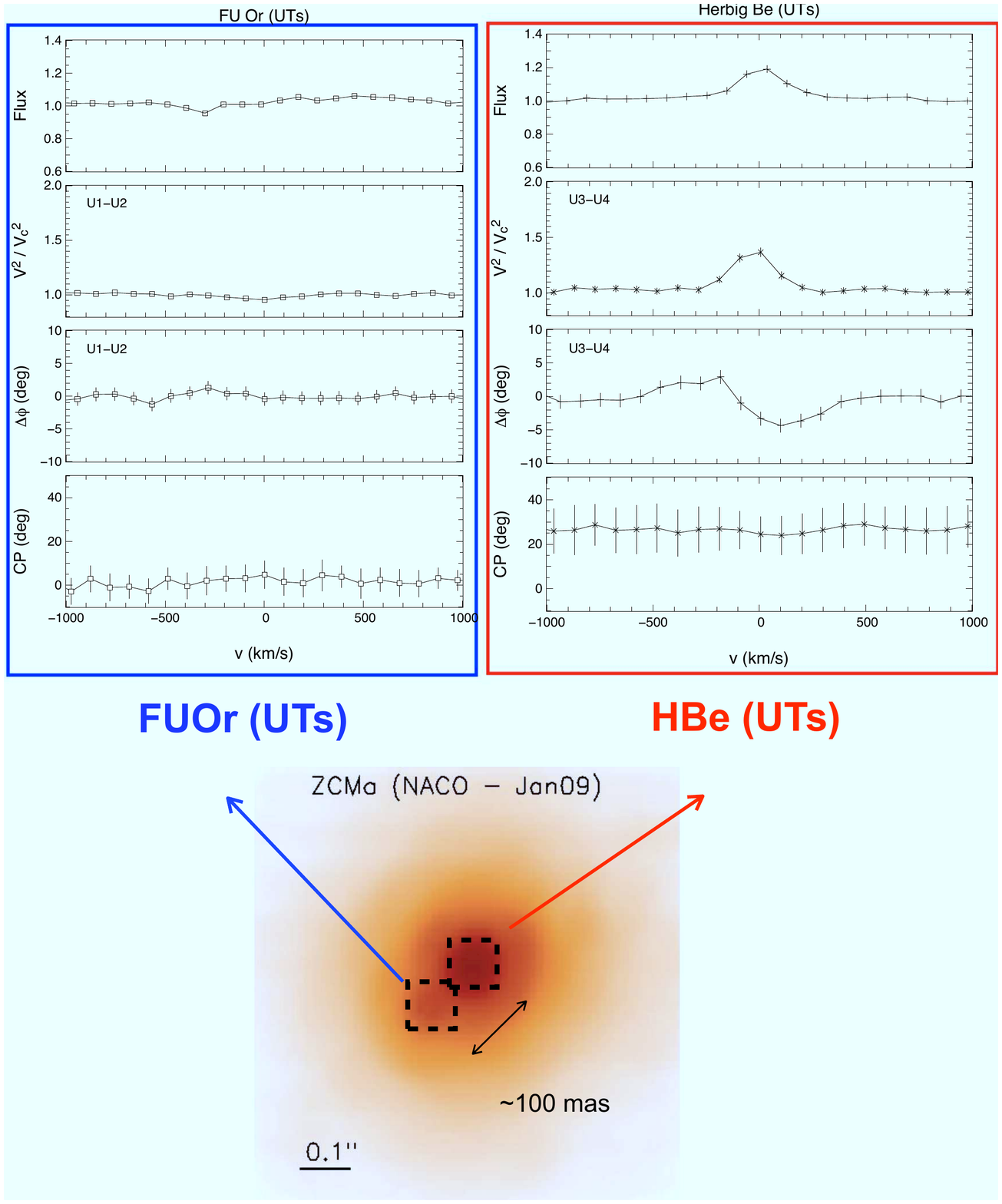}
  \caption{Interferometric measurements of the \zcma\ binary system
    with the UTs during the outburst in December~2008. Medium
    resolution spectra, spectrally dispersed squared visibilities
    normalized to the continuum ones V$_{\rm{c}}^{2}$, differential
    phases ($\Delta\Phi$), and closure phases (CP) measured for the
    FU~Or (left) and the same quantities for the Herbig~Be (right).
    The binary system as observed by the NACO instrument at the VLT
    (adaptive optics with infrared camera) is displayed at the bottom
    with identification of the sources. The square boxes represent the
    field of view of these individual measurements.}
\label{fig:zcma-mr-uts}
\end{figure}

The left and right columns of Fig.~\ref{fig:zcma-mr-uts} present examples
of the MR observations obtained with the UTs for each star during the
outburst. Each column includes a spectrum (normalized to the
continuum), squared visibilities, differential phases, and closure
phases. Since the absolute values of the visibilities measured with
the UTs are unknown, we normalized the continuum values to 1 -- even
though the emission is resolved.

For the FU~Or (left panels), within the error bars, the spectrum shows
\brg\ in neither emission nor absorption. Consequently, no change in
the visibilities or phases across the line is expected nor seen.  In
contrast, the Herbig~Be star exhibits a clear \brg\ line in emission
(right panels), although at this spectral resolution ($\Delta
v\sim$200\,km/s), the line is not spectrally resolved. The visibility
increases through the line and the differential phases produce an
S-shape variation. The closure phases differ from zero, with values of
25$^\circ \pm$12$^\circ$. The phase, line, and visibility signals are
present from $\sim$-600 to 500\,km/s, although because of the low
line-to-continuum ratio in the extended wings, the flux and
visibilities appear narrower.  Within the large errors, no variation
in the closure phases is detected across the line.

\subsection{Size of the Herbig Be system in the continuum and in \brg}
\label{sec:size-herbig-be}

\begin{figure}[t]
  \centering
  \includegraphics[height=0.65\textheight]{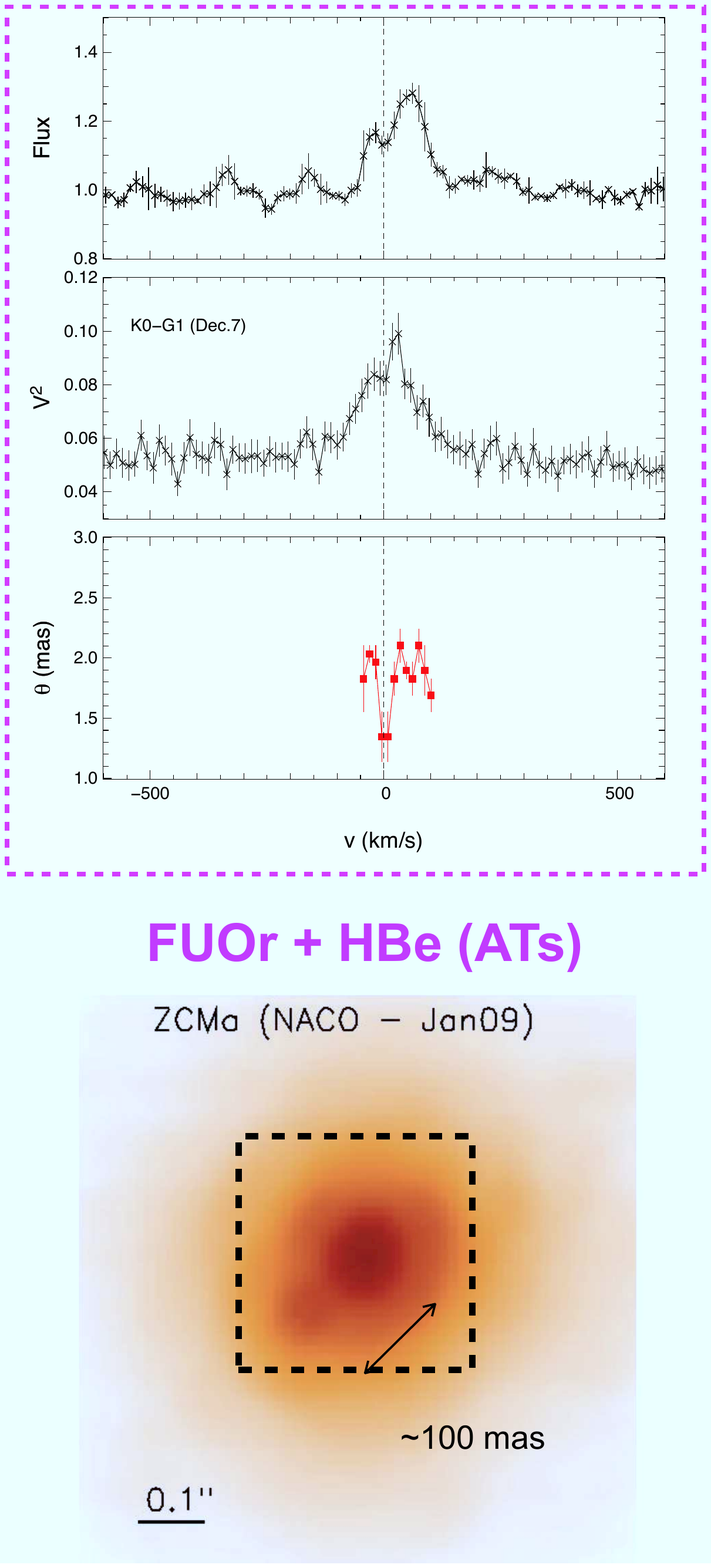}
  \caption{Interferometric measurements of the \zcma\ binary system
    with the ATs during the outburst in December~2008. High resolution
    spectra (top), spectrally dispersed absolute squared
    visibilities(middle) measured for the total system including the
    FU~Or and the Herbig~Be (right) and the corresponding equivalent
    diameters in \brg (bottom).  The binary system as observed by the
    NACO instrument at the VLT (adaptive optics with infrared camera)
    is displayed at the bottom with identification of the sources. The
    square box represents the field of view of these measurements.}
\label{fig:zcma-hr-ats}
\end{figure}

Figure \ref{fig:zcma-hr-ats} shows measurements obtained with the ATs,
\textit{i.e.}, with both stars in the field of view. In this case, the
level of continuum is determined by both stellar components.  These
panels present observations obtained in high spectral resolution
($R\sim12\,000$).  In this case, the line is spatially and spectrally
resolved ($\Delta v\sim$25\,km/s), and the spectra exhibit a clear
double-peaked and asymmetric profiles, with less emission at
blueshifted velocities. The spectral visibilities present a similar
profile.

From the visibilities, one can locate the emission at each velocity
and distinguish between various scenarios capable of producing the
line. The visibility increase within the line implies that the \brg\
emitting region is more compact than the one responsible for the
continuum. To derive the characteristic sizes of the region emitting
\brg\ only, for each spectral channel of the HR measurements, one has
to subtract the underlying continuum to first determine the visibility
of the line only\cite{weigelt07}. These estimates can only be
performed using the data gathered with the ATs, for which reliable
absolute values for the \brg\ visibilities are obtained.  Using a
model of an uniform ring, the emission in the line has a typical
extension (ring diameter) of $\sim$1.6\,mas at zero velocity, and
$\sim$2.5\,mas at higher velocities ($\sim$100\,km/s), \textit{i.e.},
from $\sim$1.5 to $\sim$2.6\,AU, depending on the distance.

As the continuum emission measured with the ATs includes
both stars, it is not direct to establish the typical size of the
Herbig~Be continuum. In contrast, the UTs data includes only one
stellar component. Although no absolute visibility values can be
obtained, size ratios between the line and the continuum can be
derived. Using the sizes previously estimated for the line from the
ATs data, typical sizes of $\sim$3.4\,mas ($\sim$3.6\,AU) for the
Herbig~Be $K$-band continuum can be determined, in agreement with the
previous estimate\cite {monnier05} ($\sim$3.9\,mas in
2004;). 

Considering a dust sublimation temperature around
1500-2000\,K\cite{pollack94}, and the stellar properties determined by
van den Ancker et al. (2004) \cite{vandenancker04}, the inner edge of
the dusty disk must be located at $\sim$4-7\,AU, in agreement with our
findings. An asymmetry in the inclined inner disk could explain the
non-zero closure phases measured at a level similar to other
Herbig~AeBe stars\cite{kraus09, benisty10}.  Such methodology is
however uncertain as it depends on the stellar contribution of the
Herbig~Be to the $K$-band continuum, that is unknown.

The \brg\ emitting region in the Herbig~Be component is therefore more
compact than the one responsible for the continuum.

\subsection{Spectro-astrometry of the \brg\ line }
\label{sec:spectro-astrometry}

\begin{figure}[t]
  \centering
  \includegraphics[width=0.5\hsize]{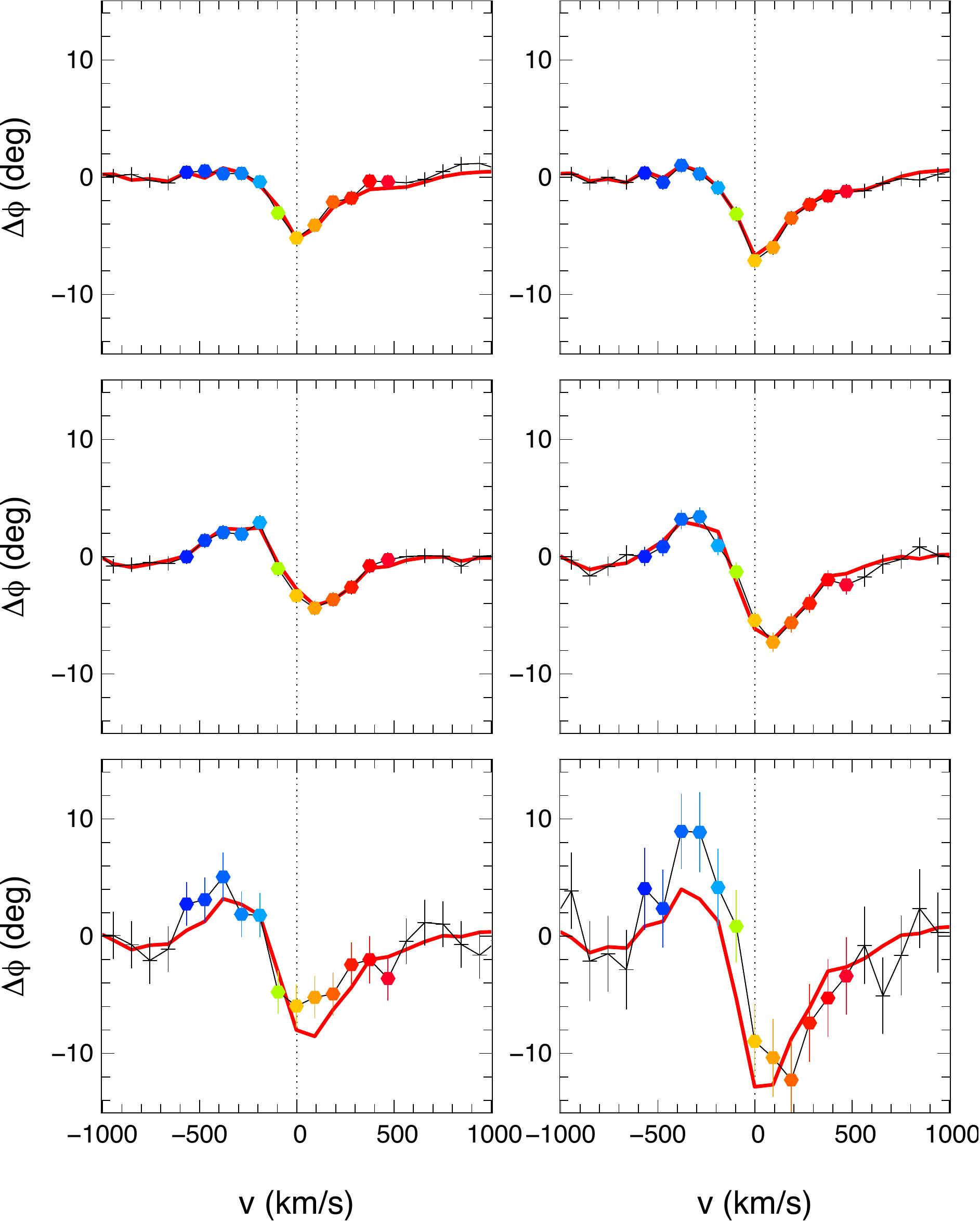}
  \caption{Differential phases measured with the UTs (black crosses
    and lines) for six different baselines.  The dots in different
    colors represent various velocity channels, from dark blue (for
    v$\sim$-600\,km/s) to dark red (for v$\sim$+500\,km/s). The red
    line is the 2D astrometric solution $\vec{p}$($\lambda$).}
  \label{fig:diffphases}
\end{figure}

The differential phases $\Delta\Phi$ which are represented in
Fig.~\ref{fig:diffphases} can be expressed in terms of photocenter
displacements \cite{lachaume03} $p$ (in arcseconds), given by
$p=-2\pi\Delta\Phi/B\lambda$, where $\lambda$ and $B$ are the
wavelength and the projected baseline length of the observations,
respectively. $p$ is the projection along the baseline direction, of
the 2D photocenter vector $\vec{p}$ in the plane of the sky
(\textit{i.e.}, of a spectro-astrometric signal).

\begin{figure}[t]
  \centering
  \includegraphics[width=0.9\hsize]{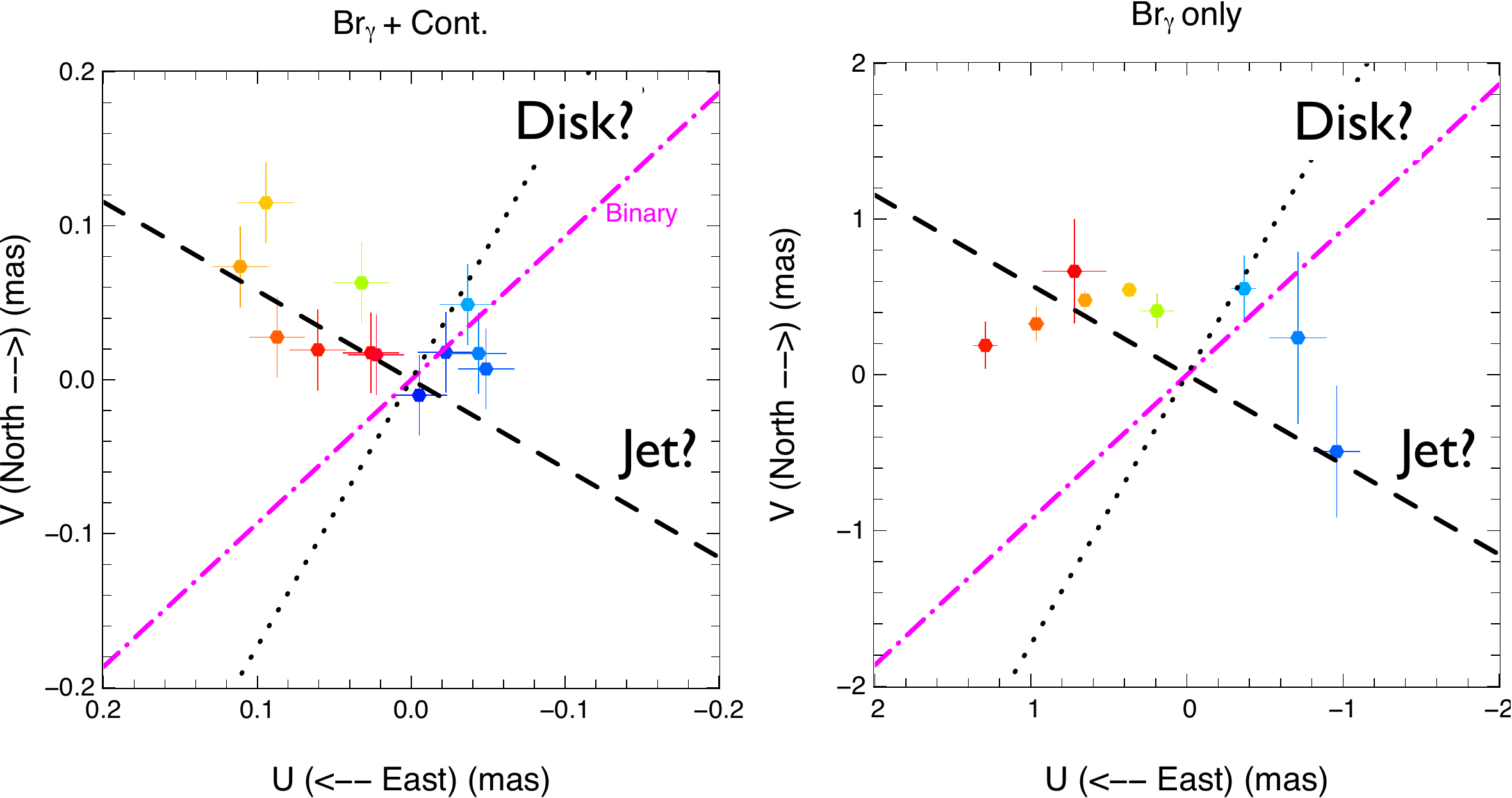}
  \caption{Left: 2D representation of $\vec{p}$($\lambda$)
  within -600 and 500 km/s. The different colors code the velocity channels as
  represented in Fig.~\ref{fig:diffphases}. The position angles of the binary
  (dashed-dotted line), of the large-scale jet (dashed line) and the
  direction perpendicular to the jet (dotted line) are overplotted. Right: same for
  $\vec{p}_{\rm{Br}_\gamma}$($\lambda$), after substraction of the
  continuum contribution. This was done within a narrower interval
  (-350 to 350\,km/s) because of the low line-to-continuum ratio in the wings.}
  \label{fig:photocenters}
\end{figure}
 
We fitted all the differential phases along the 6 available baselines
with a single vector $\vec{p}$, independently of each spectral
channel. Figure \ref{fig:diffphases} shows the differential phases and
the best solution for $\vec{p}$. The left plot of
Fig.~\ref{fig:photocenters} gives $\vec{p}$ in a 2D map of the plane
of the sky.  Clear asymmetric displacements, up to
$\sim$150\,micro-arcseconds, are observed, both at red-shifted and
blue-shifted velocities.  In this case, $\vec{p}$ accounts for the
emission of both the line and the continuum.

Subtracting the continuum contribution to determine the photocenter
displacements, $\vec{p}_{\rm{Br}_\gamma}$, due to the line only, is
difficult, as it has to be done in the complex visibility plane. We
provide such an attempt in the velocity range where the line is
clearly detected ([-350;350]\,km/s, with line-to-continuum ratio
larger than 1.05). As can be seen in the right panel of
Fig.~\ref{fig:photocenters}, the displacements are much larger (up to
$\sim$1\,mas) with the largest measured at the highest velocities, and
appear more spread. Nonetheless, the observed asymmetry is still
consistent with the closure phase measurements that show no change
through the line, within the large errors.

\section{Interpretation: Accretion or ejection signature?} 
\label{sec:discussion}

As has already been discussed in previous studies\cite{kraus08,
eisner09}, the \brg\ line could  be emitted by a variety of
 mechanisms, such as accretion of matter onto the star, in
 a gaseous disk, or in outflowing matter. The spectra obtained at
 high spectral resolution show a double-peaked 
and asymmetric profile that can be interpreted in the context of the
formation of optically thick lines in a dense environment with a
temperature gradient\cite{cesa95, kurosawa06}. 

\subsection{Infalling envelope of gas?}

Formation of the \brg\ line in an infalling envelope of gas can be
ruled out. Considering that the line excitation temperature increases
towards the star, if the line is emitted in infall of matter, or
accretion flows, the profile would be double peaked but with an
opposite asymmetry to what is observed (see
Fig.~\ref{fig:zcma-hr-ats}, \textit{i.e.} with a lower emission at
redshifted velocities\cite{hartmann94, walker94}). In addition, in
that case, the smallest extension and photocenter displacements would
be expected at the highest velocities, which is in disagreement with
our findings (see Figs.~\ref{fig:zcma-hr-ats} and
\ref{fig:photocenters}).

\subsection{Hot layers of a gaseous disk?}

The possibility that the \brg\ line forms in the hot layers of the
gaseous disk can also be ruled out. If one considers that the
circumstellar disk surrounding the Herbig~Be 
star is perpendicular to the large-scale jet at PA$\sim$240\dg, it would be
expected that the velocities projected onto the line of sight cancel
out along the  semi-minor axis, while large spectro-astrometric
displacements  are seen  along  this axis  at high  velocities
(Fig.~\ref{fig:photocenters}). Apart from this, the displacements increase with
velocity while Keplerian rotation should behave in the opposite way,
and a phase signal is measured up to high 
velocities ($\sim$500\,km/s), which are much larger than the expected Keplerian
velocities ($\sim$100-120\,km/s at $\sim$1\,AU). It therefore seems 
unlikely that the \brg\ line is emitted in the disk. 

\subsection{Bipolar wind or base of a jet?}

\begin{figure}[t]
  \centering
  \includegraphics[width=0.85\hsize]{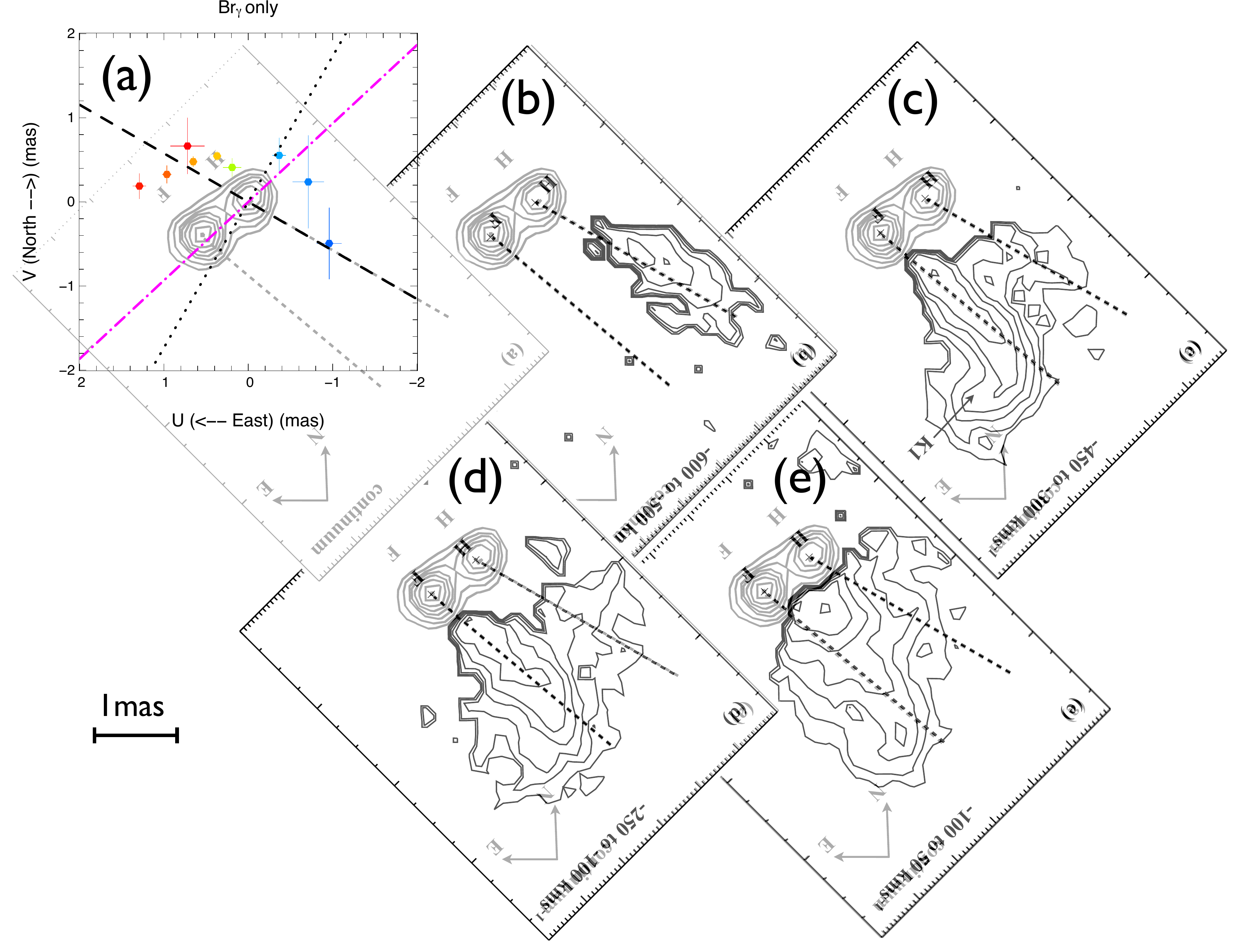}
  \caption{Is a jet the responsible for the photocenter displacement?
    Composite figure using AMBER measurements and spectro-images
    obtained in the [FeII] line on OSIRIS on Keck by Whelan et al.\ (2010)\cite{whelan10}. The continuum image of the \zcma\ has been displayed in
    grey in the background of each panels. Panel (a): displacement of
    the photocenter measured by AMBER in different channels of the
    \brg\ line corresponding to the right part of
    Fig.~\ref{fig:photocenters}. Panels (b) to (e) are images obtained
  in different velocities slot by Whelan et al.\ (2010)\cite{whelan10}. The velocities
  channels are respectively: (b) -600 to -500\,km/s, (c) -450 to
  -300\,km/s, (d) -250 to -100\,km/s and (e) -100 to 50\,km/s. The
  direction of displacement of the blueshifted photocenters seems to
  align well with the structure detected at much larger scale in the
  [FeII] wind.}
  \label{fig:jets}
\end{figure}

We consider that the most likely origin of the \brg\ emission is a
wind. Strong winds are expected to take place in massive
Herbig~Be\cite{nisini95, malbet07}, and could be responsible for the
\brg\ emission. The double peaked and asymmetric line profile is
consistent with outflowing matter emitting in optically thick
lines\cite{hartmann90}. The visibilities and the 2D maps of the
astrometric signal also support this conclusion as the blue- and
redshifted emissions are located on each side of the possible disk
position angle, with the largest \brg\ displacements and
characteristic sizes being derived at higher velocities. 

Our observations suggest that the disk is slightly inclined, to allow
both red- and blueshifted emissions to be seen. We may be seeing the
emission from a wind partly through an optically thin inner hole in
the optically thick dusty disk\cite{takami01, emma04}. Alternatively,
if the inner gaseous disk were optically thick, we may be seeing the
redshifted emission through a much smaller hole and via scattering on
the disk surface. Fig.~\ref{fig:zcma-hr-ats} suggests also that the
red lobe is larger than the blue lobe. Could this be due to the fact
that, because of the hole in the disk, we see only a smaller part of
the blue lobe on the other side of the disk? This agrees with the size
of the displacements of the photocenter.

Whether the innermost disk is optically thick or
not cannot be determined with our observations and no reliable
estimate of the mass accretion rate exists for such high mass young
stars. However, the presence of the CO overtone lines in emission
(Bonnefoy et al., in prep) is indicative of a much lower mass
accretion rate than those derived for FU~Ors ($\leq
10^{-5}$M$_{\odot}.$yr$^{-1}$)\cite{calvet91, carr89}.

At the spatial resolutions provided by the VLTI, we trace the regions
close to the inner disk hole and it is therefore unsurprising that we
could detect redshifted emission, while on scales of 10-100\,AU, the
redshifted lobe is obscured by the circumstellar disk\cite{poetzel89,
  garcia99} (see Fig.~\ref{fig:jets}; Whelan et al.\
2010\cite{whelan10}).  During this outburst, deep blueshifted
absorption was detected in the Balmer lines from zero velocity to
$\sim$700\,km/s, in addition to the absence of redshifted emission at
similar velocities\cite{monnier05} (Bouvier et~al., in~prep),
supporting our conclusion that there is a strong wind in the
Herbig~Be.

Could our new observations be tracing the base of the jet seen at
larger scale and be in fact the inner parts of the parsec scale
outflow? As shown in Fig.~\ref{fig:photocenters}, the astrometric
signal is detected at a slightly different position angle and spans a
range broader than 60$^\circ$, a value commonly assumed for the
opening angle of the jets.  At these spatial scales, it is unlikely
that the jet is already collimated. The different position angle with
respect to the larger scale jet (see Fig.~\ref{fig:jets}) could be due
to jet precession. Our observations exclude a fully spherical wind
since in that case no displacement would be expected between the
redshifted and blueshifted emission lobes. The derived
spectro-astrometric signatures favor a bipolar wind, maybe unrelated
to the jet, but can not determine whether its geometry is that of a
disk-wind or a stellar wind.

\subsection{Outburst: accretion seen through massive ejection?}
\label{sec:accretion-ejection}

After detecting the same level of optical polarisation in both
continuum and spectral lines along a position angle roughly
perpendicular to the large-scale jet, Szeifert et al.\
(2010)\cite{szeifert10} concluded that this outburst is related to a
change in the path along which the photons escape from the dust
cocoon.  The disappearance of the \brg\ emission line, with respect to
the continuum, after the outburst, suggests that its emission is
related to the outburst. A strong mass ejection event could account
for the deep blueshifted absorption features seen in the Balmer lines
that are emitted close to the star as well as for the \brg\ line
emitted in outer layers of the wind. Outside the outburst, the wind
disappears or is more likely to be maintained at a much smaller mass
loss rate. Based on these conclusions, one can speculate about the
origin of the outburst, as being driven by an event of enhanced mass
accretion, similar to the EX~Ors and FU~Ors outbursts\cite{zhu10}. In
that case, this would suggest a strong link between mass accretion and
ejection during the outburst, probably coupled with a magnetic field
as in lower-mass young
stars. 

Whether the innermost disk is optically thick or not can not
be determined with our observations and no reliable estimate of the
mass accretion rate exists for such high mass young stars. However,
the presence of the CO overtone lines in emission (Bonnefoy et al., in
prep.) suggests a much lower mass accretion rate than the ones
derived for FU~Ors.

\section{Conclusion}
\label{sec:conclusion}

We have presented spatially and spectrally resolved interferometric
observations of the $K$-band emission in the \zcma\ system. These
observations were performed during the largest photometric outburst
detected so far, that occurred in the innermost regions of the
Herbig~Be star.

We found that the \brg\ line profile, the astrometric signal, and the
characteristic sizes across the line are inconsistent with a Keplerian
disk or with infall of matter. They are, instead, evidence of a
bipolar wind seen through a disk hole, inside the dust sublimation
radius. 

The disappearance of the \brg\ emission line after the outburst
suggests that the outburst is related to a period of strong mass loss.
Based on these conclusions, we speculate that the origin of the
outburst is an event of enhanced mass accretion, and that it does not
result from a change in the system obscuration by dust. If this is
valid, our results would suggest that the link between mass accretion
and ejection as observed for quiescent T Tauri stars can also be at
play in more massive young stars, and in high-accretion states. Is
this accretion-ejection link universal, independent of the central
mass?
 
Finally, this paper illustrates the great potential of the combination
of spectro-astrometric and interferometric techniques for observing
structures on micro-arcsecond scales. It may provide strong
constraints on the mechanisms at play.

\acknowledgements

We thank the VLTI team at Paranal, as well as R. Cesaroni,
S. Antoniucci, L. Podio, P.~Stee and M.~van den Ancker for fruitful
discussions. M.B. acknowledges funding from INAF (grant ASI-INAF
I/016/07/0). 

\bibliographystyle{spiebib}
\bibliography{zcma-spie}

\begin{thebibliography}{10}

\bibitem{stahler98}
{Stahler}, S.~W., ``{Deuterium and the stellar birthline},'' {\em \apj}~{\bf
  332},  804--825 (1988).

\bibitem{bertout88}
{Bertout}, C., {Basri}, G., and {Bouvier}, J., ``{Accretion disks around T
  Tauri stars},'' {\em \apj}~{\bf 330},  350--373 (1988).

\bibitem{konigl91}
{Koenigl}, A., ``{Disk accretion onto magnetic T Tauri stars},'' {\em
  \apjl}~{\bf 370},  L39--L43 (1991).

\bibitem{calvet92}
{Calvet}, N. and {Hartmann}, L., ``{Balmer line profiles for infalling T Tauri
  envelopes},'' {\em \apj}~{\bf 386},  239--247 (1992).

\bibitem{kenyon90}
{Kenyon}, S.~J., {Hartmann}, L.~W., {Strom}, K.~M., and {Strom}, S.~E., ``{An
  IRAS survey of the Taurus-Auriga molecular cloud},'' {\em \aj}~{\bf 99},
  869--887 (1990).

\bibitem{evans09}
{Evans}, N.~J., {Dunham}, M.~M., {J{\o}rgensen}, J.~K., and {coll.}, ``{The
  Spitzer c2d Legacy Results: Star-Formation Rates and Efficiencies; Evolution
  and Lifetimes},'' {\em \apjs}~{\bf 181},  321--350 (2009).

\bibitem{2010arXiv1007.0682B}
{Benisty}, M., {Malbet}, F., {Dougados}, C., {Natta}, A., {Le Bouquin}, J.~B.,
  {Massi}, F., {Bonnefoy}, M., {Bouvier}, J., {Chauvin}, G., {Chesneau}, O.,
  {Garcia}, P.~J.~V., {Grankin}, K., {Isella}, A., {Ratzka}, T., {Tatulli}, E.,
  {Testi}, L., {Weigelt}, G., and {Whelan}, E.~T., ``{The 2008 outburst in the
  young stellar system ZCMa: I. Evidence of an enhanced bipolar wind on the
  AU-scale},'' {\em \aap}~{\bf in press},  arXiv:1007.0682 (2010).

\bibitem{grankin09}
{Grankin}, K.~N. and {Artemenko}, S.~A., ``{New Extreme Outburst of Z CMa},''
  {\em IBVS}~{\bf 5905},  1--+ (2009).

\bibitem{koresko91}
{Koresko}, C.~D., {Beckwith}, S.~V.~W., {Ghez}, A.~M., and {et al.}, ``{An
  infrared companion to Z Canis Majoris},'' {\em \aj}~{\bf 102},  2073--2078
  (1991).

\bibitem{barth94}
{Barth}, W., {Weigelt}, G., and {Zinnecker}, H., ``{Speckle masking
  observations of the young binary Z Canis Majoris},'' {\em \aap}~{\bf 291},
  500--504 (1994).

\bibitem{claria74}
{Clari{\'a}}, J.~J., ``{A study of the stellar association Canis Major OB
  1.},'' {\em \aap}~{\bf 37},  229--236 (1974).

\bibitem{kaltcheva00}
{Kaltcheva}, N.~T. and {Hilditch}, R.~W., ``{The distribution of bright OB
  stars in the Canis Major-Puppis-Vela region of the Milky Way},'' {\em
  \mnras}~{\bf 312},  753--768 (2000).

\bibitem{whitney93}
{Whitney}, B., {Clayton}, G., {Schulte-Ladbeck}, R., {Calvet}, N., {Hartmann},
  L., and {Kenyon}, S., ``{Spectrum of the ``Invisible'' Companion of Z Canis
  Majoris Revealed in Polarized Light},'' {\em \apj}~{\bf 417},  687--+ (1993).

\bibitem{alonso09}
{Alonso-Albi}, T., {Fuente}, A., {Bachiller}, R., {Neri}, R., {Planesas}, P.,
  {Testi}, L., {Bern{\'e}}, O., and {Joblin}, C., ``{Circumstellar disks around
  Herbig Be stars},'' {\em \aap}~{\bf 497},  117--136 (2009).

\bibitem{hartmann89}
{Hartmann}, L., {Kenyon}, S.~J., {Hewett}, R., {Edwards}, S., {Strom}, K.~M.,
  {Strom}, S.~E., and {Stauffer}, J.~R., ``{Pre-main-sequence disk accretion in
  Z Canis Majoris},'' {\em \apj}~{\bf 338},  1001--1010 (1989).

\bibitem{vandenancker04}
{van den Ancker}, M., {Blondel}, P., {Tjin A Djie}, H., {Grankin}, K.,
  {Ezhkova}, O., {Shevchenko}, V., {Guenther}, E., and {Acke}, B., ``{The
  stellar composition of the star formation region CMa R1 - III. A new outburst
  of the Be star component in Z CMa},'' {\em \mnras}~{\bf 349},  1516--1536
  (2004).

\bibitem{poetzel89}
{Poetzel}, R., {Mundt}, R., and {Ray}, T.~P., ``{Z CMa - A large-scale high
  velocity bipolar outflow traced by Herbig-Haro objects and a jet},'' {\em
  \aap}~{\bf 224},  L13--L16 (1989).

\bibitem{evans94}
{Evans}, N., {Balkum}, S., {Levreault}, R., and {et al.}, ``{Molecular outflows
  from FU Orionis stars},'' {\em \apj}~{\bf 424},  793--799 (1994).

\bibitem{garcia99}
{Garcia}, P.~J.~V., {Thi{\'e}baut}, E., and {Bacon}, R., ``{Spatially resolved
  spectroscopy of Z Canis Majoris components},'' {\em \aap}~{\bf 346},
  892--896 (1999).

\bibitem{monnier05}
{Monnier}, J.~D., {Millan-Gabet}, R., {Billmeier}, R., and {coll.}, ``{The
  Near-Infrared Size-Luminosity Relations for Herbig Ae/Be Disks},'' {\em
  \apj}~{\bf 624},  832--840 (2005).

\bibitem{millan06}
{Millan-Gabet}, R., {Monnier}, J.~D., {Akeson}, R.~L., and {coll.}, ``{Keck
  Interferometer Observations of FU Orionis Objects},'' {\em \apj}~{\bf 641},
  547--555 (2006).

\bibitem{szeifert10}
{Szeifert}, T., {Hubrig}, S., {Sch{\"o}ller}, M., {Sch{\"u}tz}, O., {Stelzer},
  B., and {Mikul{\'a}{\v s}ek}, Z., ``{The nature of the recent extreme
  outburst of the Herbig Be/FU Orionis binary Z Canis Majoris},'' {\em
  \aap}~{\bf 509},  L7+ (2010).

\bibitem{vlti1}
{Sch{\"o}ller}, M., ``{The Very Large Telescope Interferometer: Current
  facility and prospects},'' {\em New Astronomy Review}~{\bf 51},  628--638
  (2007).

\bibitem{petrov07}
{Petrov}, R.~G., {Malbet}, F., {Weigelt}, G., and {coll.}, ``{AMBER, the
  near-infrared spectro-interferometric three-telescope VLTI instrument},''
  {\em \aap}~{\bf 464},  1--12 (2007).

\bibitem{lebouquin08}
{Le Bouquin}, J.-B., {Bauvir}, B., {Haguenauer}, P., {Sch{\"o}ller}, M.,
  {Rantakyr{\"o}}, F., and {Menardi}, S., ``{First result with AMBER+FINITO on
  the VLTI: the high-precision angular diameter of V3879 Sagittarii},'' {\em
  \aap}~{\bf 481},  553--557 (2008).

\bibitem{tatulli07}
{Tatulli}, E., {Millour}, F., {Chelli}, A., and {coll.}, ``{Interferometric
  data reduction with AMBER/VLTI. Principle, estimators, and illustration},''
  {\em \aap}~{\bf 464},  29--42 (2007).

\bibitem{chelli09}
{Chelli}, A., {Utrera}, O.~H., and {Duvert}, G., ``{Optimised data reduction
  for the AMBER/VLTI instrument},'' {\em \aap}~{\bf 502},  705--709 (2009).

\bibitem{weigelt07}
{Weigelt}, G., {Kraus}, S., {Driebe}, T., and {coll.}, ``{Near-infrared
  interferometry of {$\eta$} Carinae with spectral resolutions of 1 500 and 12
  000 using AMBER/VLTI},'' {\em \aap}~{\bf 464},  87--106 (2007).

\bibitem{pollack94}
{Pollack}, J.~B., {Hollenbach}, D., {Beckwith}, S., {Simonelli}, D.~P.,
  {Roush}, T., and {Fong}, W., ``{Composition and radiative properties of
  grains in molecular clouds and accretion disks},'' {\em \apj}~{\bf 421},
  615--639 (1994).

\bibitem{kraus09}
{Kraus}, S., {Hofmann}, K., {Malbet}, F., {Meilland}, A., {Natta}, A.,
  {Schertl}, D., {Stee}, P., and {Weigelt}, G., ``{Revealing the sub-AU
  asymmetries of the inner dust rim in the disk around the Herbig Ae star R
  Coronae Austrinae},'' {\em \aap}~{\bf 508},  787--803 (2009).

\bibitem{benisty10}
{Benisty}, M., {Natta}, A., {Isella}, A., and {coll.}, ``{Strong near-infrared
  emission in the sub-AU disk of the Herbig Ae star HD 163296: evidence of
  refractory dust?},'' {\em \aap}~{\bf 511},  A74+ (2010).

\bibitem{lachaume03}
{Lachaume}, R., ``{On marginally resolved objects in optical interferometry},''
  {\em \aap}~{\bf 400},  795--803 (2003).

\bibitem{kraus08}
{Kraus}, S., {Hofmann}, K., {Benisty}, M., and {coll.}, ``{The origin of
  hydrogen line emission for five Herbig Ae/Be stars spatially resolved by
  VLTI/AMBER spectro-interferometry},'' {\em \aap}~{\bf 489},  1157--1173
  (2008).

\bibitem{eisner09}
{Eisner}, J.~A., {Graham}, J.~R., {Akeson}, R.~L., and {Najita}, J.,
  ``{Spatially Resolved Spectroscopy of Sub-AU-Sized Regions of T Tauri and
  Herbig Ae/Be Disks},'' {\em \apj}~{\bf 692},  309--323 (2009).

\bibitem{cesa95}
{Cesaroni}, R., ``{An analytical method for computing optically thick line
  profiles.},'' {\em \aaps}~{\bf 114},  397--+ (1995).

\bibitem{kurosawa06}
{Kurosawa}, R., {Harries}, T.~J., and {Symington}, N.~H., ``{On the formation
  of H{$\alpha$} line emission around classical T Tauri stars},'' {\em
  \mnras}~{\bf 370},  580--596 (2006).

\bibitem{hartmann94}
{Hartmann}, L., {Hewett}, R., and {Calvet}, N., ``{Magnetospheric accretion
  models for T Tauri stars. 1: Balmer line profiles without rotation},'' {\em
  \apj}~{\bf 426},  669--687 (1994).

\bibitem{walker94}
{Walker}, C.~K., {Narayanan}, G., and {Boss}, A.~P., ``{Spectroscopic
  signatures of infall in young protostellar systems},'' {\em \apj}~{\bf 431},
  767--782 (1994).

\bibitem{whelan10}
{Whelan}, E., {Dougados}, C., {Perrin}, M., {Bonnefoy}, M., {Bains}, I.,
  {Redman}, M., {Ray}, T.~P., {Bouy}, H., {Benisty}, M., {Bouvier}, J.,
  {Chauvin}, G., {Garcia}, P., {Grankvin}, K., and {Malbet}, F., ``{The 2008
  Outburst in the Young Stellar System Z CMa: The First Detection of Twin
  Jets},'' {\em \apj}~{\bf in press} (2010).

\bibitem{nisini95}
{Nisini}, B., {Milillo}, A., {Saraceno}, P., and {Vitali}, F., ``{Mass loss
  rates from HI infrared lines in Herbig Ae/Be stars.},'' {\em \aap}~{\bf 302},
   169--+ (1995).

\bibitem{malbet07}
{Malbet}, F., {Benisty}, M., {de Wit}, W., and {coll.}, ``{Disk and wind
  interaction in the young stellar object MWC 297 spatially resolved with
  AMBER/VLTI},'' {\em \aap}~{\bf 464},  43--53 (2007).

\bibitem{hartmann90}
{Hartmann}, L., {Avrett}, E.~H., {Loeser}, R., and {Calvet}, N., ``{Winds from
  T Tauri stars. I - Spherically symmetric models},'' {\em \apj}~{\bf 349},
  168--189 (1990).

\bibitem{takami01}
{Takami}, M., {Bailey}, J., and {Gledhill}, T.~M. e.~a., ``{Circumstellar
  structure of RU Lupi down to au scales},'' {\em \mnras}~{\bf 323},  177--187
  (2001).

\bibitem{emma04}
{Whelan}, E.~T., {Ray}, T.~P., and {Davis}, C.~J., ``{Paschen beta emission as
  a tracer of outflow activity from T-Tauri stars, as compared to optical
  forbidden emission},'' {\em \aap}~{\bf 417},  247--261 (2004).

\bibitem{calvet91}
{Calvet}, N., {Patino}, A., {Magris}, G.~C., and {D'Alessio}, P.,
  ``{Irradiation of accretion disks around young objects. I - Near-infrared CO
  bands},'' {\em \apj}~{\bf 380},  617--630 (1991).

\bibitem{carr89}
{Carr}, J.~S., ``{Near-infrared CO emission in young stellar objects},'' {\em
  \apj}~{\bf 345},  522--535 (1989).

\bibitem{zhu10}
{Zhu}, Z., {Hartmann}, L., {Gammie}, C.~F., {Book}, L.~G., {Simon}, J.~B., and
  {Engelhard}, E., ``{Long-term Evolution of Protostellar and Protoplanetary
  Disks. I. Outbursts},'' {\em \apj}~{\bf 713},  1134--1142 (2010).

\end{thebibliography}

\end{document}